\documentclass[12pt,a4paper]{article}

\usepackage[intlimits]{amsmath}
\usepackage{amssymb}
\usepackage{graphicx}
\usepackage[dvips,a4paper,scale={.75,.8}]{geometry}
\usepackage{mathrsfs}
\usepackage{axodraw}
\usepackage{caption}
\usepackage{subfigure}
\usepackage{enumerate}
\usepackage{makeidx}

\usepackage{cancel}

\def\s1{\hat{s}_1}
\def\T1{\hat{t}_1}
\def\t2{\hat{t}_2}
\def\U1{\hat{u}_1}
\def\u2{\hat{u}_2}

\newcommand{\comment}[1]{}

\usepackage{epsfig}

\usepackage{amssymb}

\begin{document}



\begin{center}
{\Large \bf Comparison of associated Higgs boson-radion and Higgs
boson pair production processes \hfill\\}
\end{center}
\vspace{0.5cm}


\begin{center}
{E.~Boos$^{1}$, S.~Keizerov$^{1}$, E.~Rakhmetov$^1$, K.~Svirina$^{1,2}$\\
\hfill\\
{\small \it $^1$Skobeltsyn Institute of Nuclear Physics, Lomonosov Moscow State University}\\
{\small \it  Leninskie Gory, 119991, Moscow, Russia} \\
 {\small \it $^2$Faculty of Physics, Lomonosov Moscow State University,
 Leninskie Gory}\\
{\small \it 119991, Moscow, Russia}  }
\end{center}





\begin{abstract}

Many models, in particular, the brane-world models with two
branes, predict the existence of the scalar radion, whose mass can
be somewhat smaller than those of all the Kaluza-Klein modes of
the graviton and Standard Model (SM) particles. Due to its origin the
radion interacts with the trace of the energy-momentum tensor of the SM.
The fermion part of the radion interaction Lagrangian is different
from that for the SM Higgs boson due to the presence
of additional terms playing a role for off-shell fermions.
It was shown previously \cite{Boos:2014} that for the case of the single
radion and single Higgs boson production processes in association with an
arbitrary number of SM gauge bosons all the contributions to the perturbative
amplitudes appearing due to these additional terms were cancelled out
making the processes similar up to a replacement of masses and overall
coupling constants. For the case of the associated Higgs boson-radion and the
Higgs boson pair production processes involving the SM gauge bosons the
similarity property also takes place. However a detailed consideration
shows that in this case it is not enough to replace simply the masses and the
constants ($m_h \rightarrow m_r$ and $v \rightarrow \Lambda_r$). One
should also rescale the triple Higgs coupling by the factor $\xi \equiv 1
+ \frac{m_r^2 - m_h^2}{3m_h^2}$.

\end{abstract}

\comment{
\begin{keyword} Radion, Higgs boson, brane-world models.


\end{keyword}
}



\section{Introduction}
One of the characteristic features of brane world models, in
particular, of the Randall-Sundrum (RS) model
\cite{Randall:1999ee} with a stabilization  of the extra space
dimension \cite{Goldberger:1999un, wolfe}, is the existence of the
radion \cite{Goldberger:1999un, Csaki:1999mp, Charmousis:1999rg}
-- the lowest Kaluza-Klein (KK) mode of the five-dimensional
scalar field appearing from the fluctuations of the metric
component corresponding to the extra dimension. The radion might
be significantly lighter than the other KK modes
\cite{Csaki:2000zn, Boos:2004uc, Boos:2005dc}, and therefore it is
of a special interest for collider phenomenology (see, e.g.,
\cite{Giudice:2000av} - \cite{Boos:2015xma}).


The radion couples to the trace of the energy-momentum tensor of the SM, so the interaction Lagrangian has the following form
\cite{Goldberger:1999un}

\begin{equation}
 L = - \frac{r(x)}{\Lambda_r} T_{\mu}^{\mu},
 \label{L1}
\end{equation}
where ${\Lambda_r}$ is a dimensional scale parameter, $r(x)$ 
stands for the radion field and $T_{\mu}^{\mu}$ is the trace of
the SM energy-momentum tensor. In most of the studies the latter
is taken at the lowest order in the SM couplings and the
fields are supposed to be on the mass shell. Here we consider the
additional terms which come into play for the case of off-shell
fermions, so the SM energy-momentum tensor has the following form
\cite{Boos:2014}:
\begin{equation}
\begin{split}
T_{\mu}^{\mu}=\, &
\frac{\beta(g_s)}{2g_s}G^{ab}_{\rho\sigma}G_{ab}^{\rho\sigma} +
\frac{\beta(e)}{2e}F_{\rho\sigma}F^{\rho\sigma}
+ \sum _{f}\left[\frac{3i}{2} \left(\left( D_\mu\bar{f} \right)\gamma ^{\mu } f
- \bar{f}\gamma^{\mu} \left( D_\mu f \right) \right) +4m_{f} \bar{f}f \left(1+\frac{h}{v} \right) \right] \\
&-\left(\partial _{\mu }
h\right)\left(\partial ^{\mu } h\right)+2m_{h}^{2} h^{2}
\left(1+\frac{h}{2v} \right)^{2}
- \left( 2m_{W}^{2} W_{\mu }^{+} W^{\mu \, -}
+m_{Z}^{2} Z_{\mu } Z^{\mu }\right) \left(1+\frac{h}{v} \right)^{2}
 \label{Tr1}
\end{split}
\end{equation}
where the first two terms correspond to the conformal anomaly of
massless gluon and photon fields, $\beta(g_s)$, $\beta(e)$ are the
QCD and QED $\beta$-functions respectively, $h$, $W^\pm$ and $Z$
are the SM Higgs, W- and Z-boson fields, $D_\mu$ is the Standard
Model covariant derivative and the summation here is carried out over
all the Standard Model fermions.

In the case of on-shell fermions the fermion part of the
Lagrangian (\ref{L1}) is the same as for the Higgs boson (with the
replacement ${\Lambda_r} \rightarrow v$), but for off-shell
fermions  additional terms need to be taken into consideration.
These terms in the Lagrangian give additional momentum
depending contributions to the fermion-antifermion-radion
interaction vertices and new
fermion-antifermion-gauge~boson-radion vertices which modify
 the radion production and decay processes making them
potentially different from the same processes with the Higgs
boson. In paper \cite{Boos:2014} it was shown  that all the
additional contributions as compared to the Higgs boson case are
canceled out in the sum of amplitudes for the single radion
production in association with an arbitrary number of any SM
vector gauge bosons and there remain Higgs-like terms only. This
property follows from the structure of any massive fermion current
emitting the radion and gauge bosons both for the case of real
and/or virtual emmited particles as well as for the case of boson
and fermion loops \cite{Boos:2014}.

In the present paper we consider the associated Higgs boson-radion
 production  in comparison to the Higgs boson pair
production processes as a continuation of our previous study
\cite{Boos:2014}. We demonstrate that in the case of the
associated Higgs boson-radion production the similarity property
is more involved. It is not enough to perform the replacement of
two constants ($m_r \rightarrow m_h$ and $\Lambda_r \rightarrow
v$) for getting the amplitude involving the radion from the
corresponding  amplitude for the Higgs boson. It is explicitly 
demonstrated that the amplitude of the  associated
production of the Higgs boson and the radion and an arbitrary
number of gauge bosons can be obtained from the  corresponding
amplitude involving the Higgs boson pair by the replacement of the
Higgs and the radion masses, the constant $\Lambda_r$ and the
Higgs vacuum expectation value $v$ and in addition by a rescaling
the triple Higgs coupling by a certain factor. As in our previous
study we do not consider the well-known differences between
the Higgs and the radion processes caused by the conformal
anomalies.

The investigation of double Higgs boson production is an important
task for experimental measurements of the Higgs field
potential profile. This problem is rather tricky even in the high
luminosity mode of the LHC, being one of the key arguments for the
ILC construction.  However, if one of the multidimensional
brane world scenarios occurs in nature, the presence of the
radion can further complicate the problem of the Higgs potential
research due to the similarity of the Higgs boson and the radion
properties.

\section{Associated Higgs boson-radion production in fermion-antifermion annihiation}
\label{sec:Associated_ee} 
\begin{figure*}[!h!]
\begin{center}
\hspace{-6mm}
\includegraphics[height=3cm]{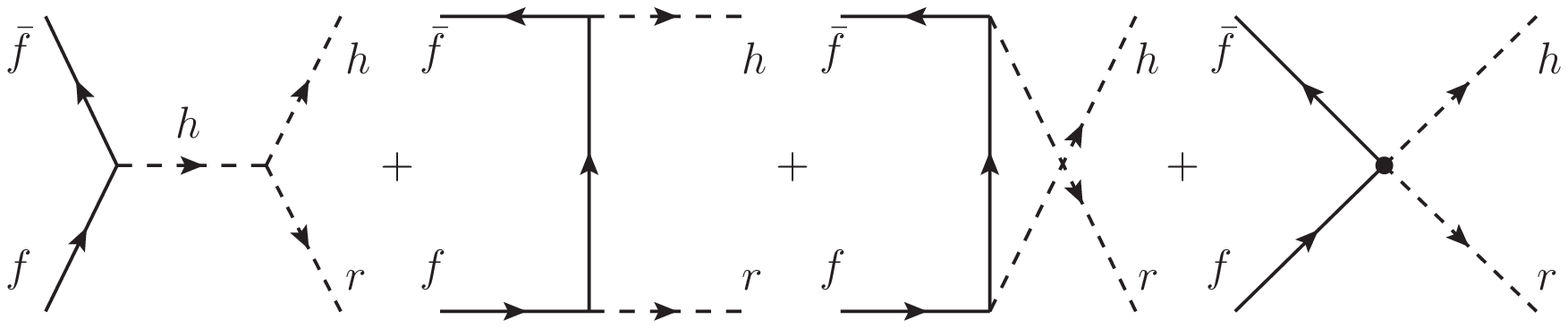}~~~
\vspace*{-7mm}
\end{center}
\caption[]{Feynman diagrams contributing to the asscociated Higgs boson-radion production in fermion-antifermion annihilation.} \label{fig:Double-ee-diag}
\vspace*{-2mm}
\end{figure*}

Let us first consider the
associated Higgs boson-radion production in fermion-antifermion
annihilation (Fig.\ref{fig:Double-ee-diag}).

The corresponding contributions to the amplitude simplified
using the Dirac equation and the identity $(\cancel{k} - m_f)\,
\frac{\cancel{k} + m_f}{k^2 - m_f^2}=1$  have the following form
\begin{equation}
M^{(rh)}_1 = \bar{v}^{r}(p_1) \, \frac{-i\, m_f}{v}\, u^{s}(p_2)\, \frac{i}{k_h^2 - m_h^2}\, \frac{-i}{\Lambda_r}\, (-2k_h\, p_h +4m_h^2)\, r(p_r)\, h(p_h)
\label{Mrh_1}
\end{equation}

\begin{equation}
\begin{split}
M^{(rh)}_2 = \bar{v}^{r}(p_1) \, \frac{-i}{\Lambda_r}\, \left\{ \frac{3}{2}\, [(\cancel{p}_1 + m_f) - (\cancel{k} - m_f)] + m_f \right\}\, r(p_r)\, i\, \frac{\cancel{k}
+ m_f}{k^2 - m_f^2}\, i\, \frac{-m_f}{v}\, u^{s}(p_2)\, h(p_h) \\
= -i\, \bar{v}^{r}(p_1) \, \frac{1}{\Lambda_r}\, \frac{m_f}{v}\, \left\{ -\frac{3}{2}\, + m_f\, \frac{\cancel{k} + m_f}{k^2 - m_f^2}\right\} \, r(p_r)\, u^{s}(p_2)\, h(p_h)
\label{Mrh_2}
\end{split}
\end{equation}
\begin{equation}
\begin{split}
M^{(rh)}_3 = \bar{v}^{r}(p_1) \, i\, \frac{-m_f}{v}\, h(p_h)\, i\, \frac{\cancel{k}^\prime + m_f}{{k^\prime}^2 - m_f^2}\,
\frac{-i}{\Lambda_r}\, \left\{ \frac{3}{2}\, [(-\cancel{k}^\prime + m_f) - (\cancel{p}_2\, - m_f)] + m_f \right\} \\ r(p_r)\, u^{s}(p_2) \\
= -i\, \bar{v}^{r}(p_1) \, \frac{1}{\Lambda_r}\, \frac{m_f}{v}\, \left\{ -\frac{3}{2}\, + m_f\, \frac{\cancel{k}^\prime + m_f}{{k^\prime}^2 - m_f^2}\right\} \, r(p_r)\, u^{s}(p_2)\, h(p_h)
\label{Mrh_3}
\end{split}
\end{equation}
\begin{equation}
M^{(rh)}_4 = \bar{v}^{r}(p_1) \, \frac{-i}{\Lambda_r}\, \frac{4 m_f}{v}\, h(p_r)\, r(p_r)\, u^{s}(p_2)\, =
 -i\, \bar{v}^{r}(p_1) \, \frac{1}{\Lambda_r}\, \frac{m_f}{v}\, \left\{ 4\right\} \, r(p_r)\, u^{s}(p_2)\, h(p_h)
\label{Mrh_4}
\end{equation}

For clarity one can modify (\ref{Mrh_1}) using the simple kinematics relation:
\begin{equation}
k_h = p_h + p_r \quad \Rightarrow \quad p_h = k_h - p_r
\end{equation}
\begin{equation}
2k_h\, p_h = 2(p_h + p_r)\, p_h = (p_h + p_r)^2 + p_h^2 - p_r^2 = k_h^2 + p_h^2 - p_r^2 = k_h^2 + m_h^2 - m_r^2
\end{equation}
therefore
\begin{equation}
\frac{-(2k_h\, p_h) +4m_h^2}{k_h^2 - m_h^2} = \frac{-k_h^2 - m_h^2 + m_r^2 + 4m_h^2}{k_h^2 - m_h^2} = -1 + \frac{m_r^2 + 2m_h^2}{k_h^2 - m_h^2}
\end{equation}

So (\ref{Mrh_1}) can be rewritten as follows
\begin{equation}
M^{(rh)}_1 = -i\, \frac{m_f}{\Lambda_r\, v}\, \bar{v}^{r}(p_1)\, r(p_r)\, h(p_h) \, \left\{ -1 + \frac{m_r^2 + 2m_h^2}{k_h^2 - m_h^2} \right\}\, u^s(p_2)
\label{Mrh_1a}
\end{equation}

It is easy to put (\ref{Mrh_2}), (\ref{Mrh_3}), (\ref{Mrh_4}) and (\ref{Mrh_1a}) together and write down the total amplitude for the $f\bar{f} \rightarrow rh$ process
\begin{equation}
\begin{split}
M^{(rh)}_{tot} = -i\, \frac{m_f}{\Lambda_r\, v}\, r(p_r)\, h(p_h) \, \bar{v}^{r}(p_1)\,
\left\{ m_f\, \frac{\cancel{k} + m_f}{k^2 - m_f^2} +
m_f\, \frac{\cancel{k}^\prime + m_f}{{k^\prime}^2 - m_f^2}
-\frac{3}{2} -\frac{3}{2} +4 -1 \right. \\
\left. + \frac{m_r^2 + 2m_h^2}{k_h^2 - m_h^2} \right\} \, u^s(p_2)
\end{split}
\label{Mrh}
\end{equation}

Now let us write down the contributions to the amplitude of
the double Higgs boson production process $f\bar{f} \rightarrow
hh$ (Fig.\ref{fig:Double-Higgs-tree}) and compare the results:
\begin{figure*}[!h!]
\begin{center}
\hspace{-6mm}
\includegraphics[height=3cm]{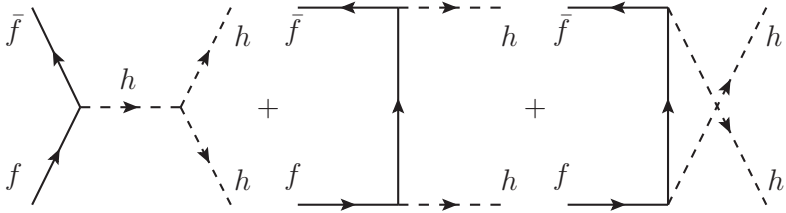}~~~
\vspace*{-7mm}
\end{center}
\caption[]{Feynman diagrams contributing to the double Higgs boson production in fermion-antifermion annihilation.} \label{fig:Double-Higgs-tree}
\vspace*{-2mm}
\end{figure*}

\begin{equation}
\begin{split}
M^{(hh)}_1 = \bar{v}^{r}(p_1) \, i \frac{-m_f}{v}\, u^s(p_2)\, \frac{i}{k_h^2 - m_h^2}\, \frac{-i}{v}3m_h^2\, h(p_{h_1})\, h(p_{h_2})\\
= -i\frac{m_f}{v^2}\,h(p_{h_1})\, h(p_{h_2})\,\bar{v}^{r}(p_1)\,\left\{\frac{3m_h^2}{k_h^2 - m_h^2} \right\}\, u^{s}(p_2)
\label{Mhh_1}
\end{split}
\end{equation}
\begin{equation}
\begin{split}
M^{(hh)}_2 = \bar{v}^{r}(p_1) \, i \frac{-m_f}{v}\, h(p_{h_1})\, i \frac{\cancel{k} + m_f}{k^2 - m_f^2}\, i\frac{-m_f}{v}\, h(p_{h_2}) \, u^s(p_2)\\
= -i\frac{m^2_f}{v^2}\,h(p_{h_1})\, h(p_{h_2})\, \bar{v}^{r}(p_1) \, \left\{\frac{\cancel{k} + m_f}{k^2 - m_f^2} \right\} \, u^{s}(p_2)
\label{Mhh_2}
\end{split}
\end{equation}
\begin{equation}
\begin{split}
M^{(hh)}_3 = \bar{v}^{r}(p_1) \, i \frac{-m_f}{v}\, h(p_{h_2})\, i \frac{\cancel{k^\prime} + m_f}{{k^\prime}^2 - m_f^2}\, i\frac{-m_f}{v}\, h(p_{h_1}) \, u^s(p_2)\\
= -i\frac{m^2_f}{v^2}\,h(p_{h_1})\, h(p_{h_2})\, \bar{v}^{r}(p_1) \, \left\{\frac{\cancel{k^\prime} + m_f}{{k^\prime}^2 - m_f^2} \right\} \, u^{s}(p_2)
\label{Mhh_3}
\end{split}
\end{equation}
Notice that in this case there is no contribution like (\ref{Mrh_4}).

Thus, the total amplitude of the double Higgs production
process yields
\begin{equation}
M^{(hh)}_{tot} = -i\, \frac{m_f}{v^2}\, h(p_{h1})\, h(p_{h2}) \, \bar{v}^{r}(p_1)\, \left\{ m_f\, \frac{\cancel{k} + m_f}{k^2 - m_f^2} +
m_f\, \frac{\cancel{k}^\prime + m_f}{{k^\prime}^2 - m_f^2} + \frac{3m_h^2}{k_h^2 - m_h^2} \right\}\, u^s(p_2)
\label{Mhh}
\end{equation}

Finally one can compare (\ref{Mrh}) and (\ref{Mhh}) and see the
explicit cancellation of all the contributions that make the
difference between the associated Higgs boson-radion and the double
Higgs boson production. 

In other words, (\ref{Mrh}) can be written in terms of (\ref{Mhh}) in the following way
\begin{equation}
M^{(rh)}_{tot} \sim \bar{v}^{r}(p_1)\,\left\{ m_f\, \frac{\cancel{k} + m_f}{k^2 - m_f^2} +
m_f\, \frac{\cancel{k}^\prime + m_f}{{k^\prime}^2 - m_f^2} + \xi \frac{3m_h^2}{k_h^2 - m_h^2} \right\}\, u^s(p_2)
\label{Mrh_xi}
\end{equation}
i.e, the expressions for the total amplitudes (\ref{Mrh}) and (\ref{Mhh}) coincide up to the replacements of
the masses $m_r \rightarrow m_h$ and the denominators of the
coupling constants $\Lambda_r \rightarrow v$ and to the renormalization of the triple Higgs coupling by the factor $\xi$, where
$$\xi \equiv 1 + \frac{m_r^2 - m_h^2}{3m_h^2}.$$

\section{Associated Higgs boson-radion production in $gg$ fusion}
\label{sec:Associated_gg} As another example let us compare two
processes involving gluons: the associated Higgs boson-radion
production ($gg \rightarrow rh$) and the double Higgs boson
production ($gg \rightarrow hh$), the corresponding diagrams are
shown below.

\begin{figure*}[!h!]
\begin{center}
\hspace{-6mm}
\includegraphics[height=8cm]{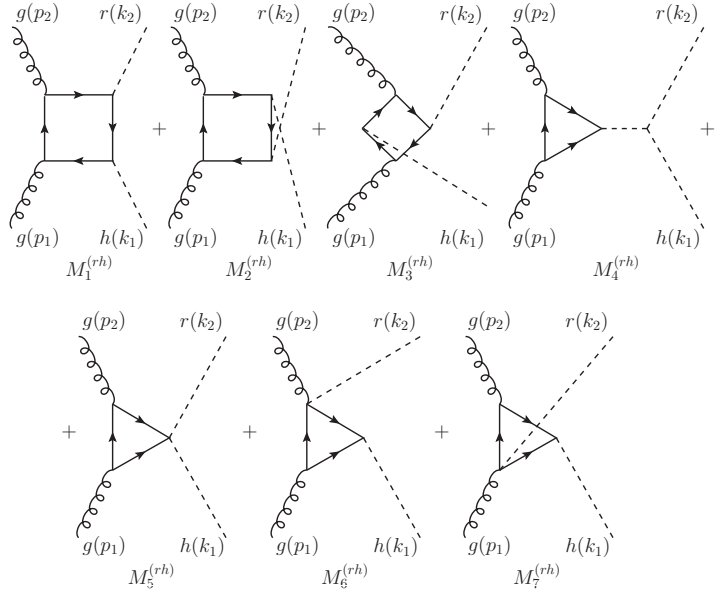}~~~
\vspace*{-5mm}
\end{center}
\caption[]{Feynman diagrams contributing to the associated Higgs boson-radion
production ($gg \rightarrow rh$).} \label{fig:Double-gg-diag1}
\vspace*{-2mm}
\end{figure*}

\begin{figure*}[!h!]
\begin{center}
\hspace{-6mm}
\includegraphics[height=4cm]{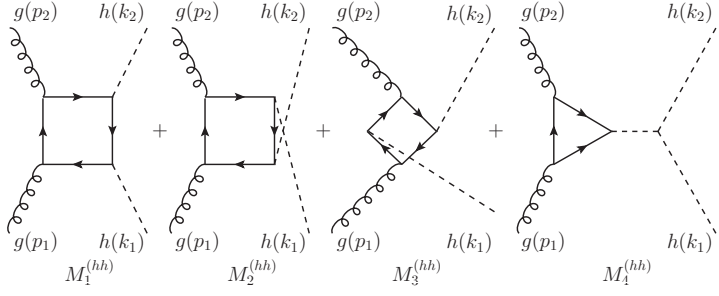}~~~
\vspace*{-5mm}
\end{center}
\caption[]{Feynman diagrams contributing to the double Higgs boson production ($gg \rightarrow hh$).} \label{fig:Double-gg-diag2}
\vspace*{-2mm}
\end{figure*}

One can see that the first four diagrams for the process of
the radion production (Fig.\ref{fig:Double-gg-diag1}) are similar
to those which appear in the SM (Fig.\ref{fig:Double-gg-diag2}).
But the other three diagrams contain the Higgs
boson-fermion-fermion-radion vertex which does not exist in the
SM.


One can notice that all the contributions to the amplitudes of these processes have the following similar structure

\begin{equation}
M^{(rh)}_i = \frac{g_c^2}{v\, \Lambda_r}\, \epsilon (p_1)_\mu\, \epsilon (p_2)_\nu\, h(k_1)\, r(k_2)\, \int \frac{d^d l}{(2\pi)^d}\, X_i^{\mu\nu}(p_1,p_2,k_1,k_2)
\end{equation}
for $gg \rightarrow rh$, where $i=1,2,...,7$;
\begin{equation}
M^{(hh)}_i = \frac{g_c^2}{v^2}\, \epsilon (p_1)_\mu\, \epsilon (p_2)_\nu\, h(k_1)\, h(k_2)\, \int \frac{d^d l}{(2\pi)^d}\, Y_i^{\mu\nu}(p_1,p_2,k_1,k_2)
\end{equation}
for $gg \rightarrow hh$, where $i=1,2,3,4$;\\
where
\begin{align*}
    &X_1^{\mu\nu} \equiv Sp[\gamma^\mu\, S_1\, \gamma^\nu\, S_2\, \Gamma_{2,3}\, S_3\, (-m)\, S_4], & Y_1^{\mu\nu} \equiv Sp[\gamma^\mu\, S_1\, \gamma^\nu\, S_2\, (-m)\, S_3\, (-m)\, S_4], \\
    &X_2^{\mu\nu} \equiv Sp[\gamma^\mu\, S_1\, \gamma^\nu\, S_2\, (-m)\, S_5 \Gamma_{5,4}\, S_4], & Y_2^{\mu\nu} \equiv Sp[\gamma^\mu\, S_1\, \gamma^\nu\, S_2\, (-m)\, S_5 (-m)\, S_4], \\
    &X_3^{\mu\nu} \equiv Sp[\gamma^\mu\, S_1\, (-m)\, S_6 \gamma^\nu\, S_5\, \Gamma_{5,4}\, S_4], & Y_3^{\mu\nu} \equiv Sp[\gamma^\mu\, S_1\, (-m)\, S_6 \gamma^\nu\, S_5\, (-m)\, S_4], \\
    &X_4^{\mu\nu} \equiv Sp[\gamma^\mu\, S_1\, \gamma^\nu\, S_2\, (-m)\, S_4]\,D\, \Gamma^\prime, & Y_4^{\mu\nu}  \equiv Sp[\gamma^\mu\, S_1\, \gamma^\nu\, S_2\, (-m)\, S_4]\,D\,\Gamma^\prime, \\
    &X_5^{\mu\nu} \equiv Sp[\gamma^\mu\, S_1\, \gamma^\nu\, S_2\, (+4m)\, S_4], \\
    &X_6^{\mu\nu} \equiv Sp[\gamma^\mu\, S_1\, (-3\gamma^\nu)\, S_3\, (-m)\, S_4], \\
    &X_7^{\mu\nu} \equiv Sp[(-3\gamma^\mu)\, S_1\, \gamma^\nu \, S_2\, (-m)\, S_5],
\end{align*}

and $\Gamma_{2,3}$, $\Gamma_{5,4}$, $\Gamma^\prime$ have the following form
\begin{equation}
\Gamma_{2,3} = \frac{3}{2}S_2^{-1} + \frac{3}{2}S_3^{-1} - m
\label{gamma23}
\end{equation}
\begin{equation}
\Gamma_{5,4} = \frac{3}{2}S_5^{-1} + \frac{3}{2}S_4^{-1} - m
\label{gamma54}
\end{equation}
\begin{equation}
\Gamma^\prime =
  \begin{cases}
   -3m_h^2, \\
   2 \left\{ (k_1+k_2)_\mu k_1^\mu - 2m_h^2 \right\}
  \end{cases}
\label{gammaprime}
\end{equation}
The first line in (\ref{gammaprime}) corresponds to the Higgs boson-fermion-fermion vertex in the $gg \rightarrow hh$ process and the second line -- to the Higgs boson-radion-fermion-fermion vertex in the $gg \rightarrow rh$ process;
\begin{equation}
S_j^{-1} = (\cancel{l} - \cancel{q}_j) - m, \quad D^{-1} = (k_1+k_2)^2 - m^2,
\label{D_prop}
\end{equation}
$$q_1=0,\, q_2=-p_2,\, q_3=-p_2+k_2,\, q_4=-p_2+k_2+k_1,\, q_5=-p_2+k_1,\, q_6=k_1.$$

\begin{figure*}[!h!]
\begin{center}
\hspace{-6mm}
\includegraphics[height=4cm]{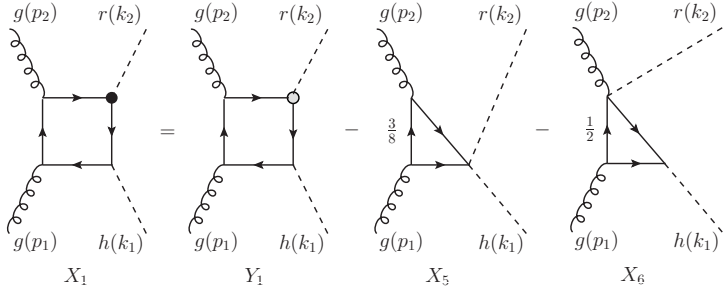}~~~
\vspace*{-3mm}
\end{center}
\caption[]{Reduction of a box diagram with the radion to a linear combination of one box diagram with the vertex such as that of the Higgs boson (the empty point) and two triangle diagrams.} \label{fig:redu_loop}
\vspace*{-2mm}
\end{figure*}

Let us notice that the radion-fermion-fermion vertex
$\Gamma_{i,j}$ contains the inverse propagators $S_i^{-1}$
and $S_j^{-1}$. In the expressions $X_1^{\mu\nu}$, $X_2^{\mu\nu}$
and $X_3^{\mu\nu}$ for the box diagrams this vertex is surrounded
by the propagators $S_i$ and $S_j$. This gives a reduction of a
box diagram with the radion to a linear combination of two
triangle diagrams and one box diagram with the vertex such as that
of the Higgs boson as it is demonstrated in
Fig.\ref{fig:redu_loop}.

\begin{figure*}[!h!]
\begin{center}
\hspace{-6mm}
\includegraphics[height=3cm]{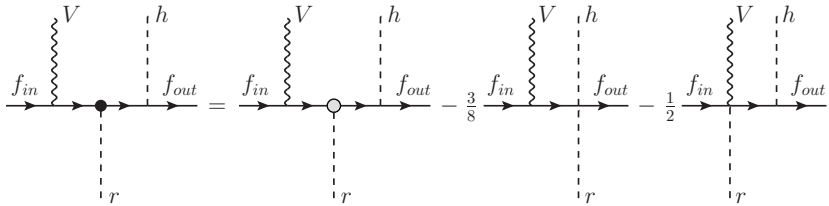}~~~
\vspace*{-3mm}
\end{center}
\caption[]{Fermion current with emission of the radion, a gauge boson and the Higgs boson expressed through the term with the Higgs-like vertex (the empty point) and two terms with four-point radion-boson vertices with corresponding numerical factors.} \label{fig:redu_tree}
\vspace*{-2mm}
\end{figure*}

This can  be easily understood with the help of the
tree-level illustration of a fermion current with emission of the
radion, a gauge boson and the Higgs boson
(Fig.\ref{fig:redu_tree}). The product of the
radion-fermion-fermion vertex (the black point in
Fig.\ref{fig:redu_tree}) and two propagators leads to three terms
respectively:
$$S_i\, \Gamma_{i,j}\, S_j = S_i\, \left\{ - m + \frac{3}{2}S_j^{-1} + \frac{3}{2}S_i^{-1} \right\} S_j$$
$$=- S_i\, m\, S_j + \frac{3}{2}\,S_i + \frac{3}{2}\,S_j$$
a Higgs-like term with the vertex proportional to the fermion mass (the empty point in Fig.\ref{fig:redu_tree}), a term with the $S_j$ propagator being dropped out, i.e., with the radion and the Higgs boson emission from the same point, and a term with the $S_i$ propagator being dropped out, with the radion and the gauge boson emission from the same point.
In this way we get a reduction of each box diagram with the radion to a sum of other contributions.

One can substitute (\ref{gamma23})--(\ref{gammaprime}) explicitly
into the expressions $X_i^{\mu\nu}$ and open the brackets in order
to get a representation of $X_1^{\mu\nu}$, $X_2^{\mu\nu}$ and
$X_3^{\mu\nu}$ as sums of other $X_i^{\mu\nu}$ and $Y_i^{\mu\nu}$
contributions with the corresponding numerical factors

\begin{equation}
X_1^{\mu\nu} = Y_1^{\mu\nu} - \frac{3}{8}X_5^{\mu\nu} - \frac{1}{2}X_6^{\mu\nu}\\
\label{xviay1}
\end{equation}
\begin{equation}
X_2^{\mu\nu} = Y_2^{\mu\nu} - \frac{3}{8}X_5^{\mu\nu} - \frac{1}{2}X_7^{\mu\nu}\\
\label{xviay2}
\end{equation}
\begin{equation}
X_3^{\mu\nu} = Y_3^{\mu\nu} - \frac{1}{2}X_6^{\mu\nu} - \frac{1}{2}X_7^{\mu\nu}
\label{xviay3}
\end{equation}

The term $X_4^{\mu\nu}$ can be represented as a combination of
$X_i^{\mu\nu}$ and $Y_i^{\mu\nu}$ contributions as follows
\begin{equation}
X_4^{\mu\nu} = -\frac{1}{4}X_5^{\mu\nu} + \frac{m_r^2 + 2m_h^2}{(k_1 + k_2)^2 - m_h^2}Y_4^{\mu\nu}
\end{equation}
by transforming the Higgs boson-radion vertex in the following way
\begin{equation}
\begin{split}
\Gamma^\prime = 2 \left\{ (k_1 + k_2)_\mu\, k_1^\mu - 2m_h^2 \right\} = (k_1 + k_2)^2 + k_1^2 - k_2^2 - 4m_h^2 \\
=[(k_1 + k_2)^2 - m_h^2] - m_r^2 - 2m_h^2
\end{split}
\end{equation}
and multiplying it by  the propagator $D$ (\ref{D_prop})
\begin{equation}
D\,\Gamma^\prime = 1 - \frac{m_r^2 + 2m_h^2}{(k_1 + k_2)^2 - m_h^2}
\end{equation}

Finally the sum of all $X_i^{\mu\nu}$ for the $gg \rightarrow rh$
process can be written in terms of $Y_i^{\mu\nu}$ and the
parameter $\xi$ again as it has been done in the previous example
(see (\ref{Mrh_xi}))
\begin{equation}
\sum_{i=1}^{7} X_i^{\mu\nu} = Y_1^{\mu\nu} + Y_2^{\mu\nu} + Y_3^{\mu\nu} + \xi \, Y_4^{\mu\nu}
\end{equation}

which looks very similar to the expression for the $gg \rightarrow hh$ process
\begin{equation}
\sum_{i=1}^{4} Y_i^{\mu\nu} = Y_1^{\mu\nu} + Y_2^{\mu\nu} + Y_3^{\mu\nu} + Y_4^{\mu\nu}
\end{equation}

Thus, once again we see that the amplitudes for these two processes coincide up to the parameter $\xi$ and to the replacements of the masses and the denominators of the coupling constants.

One can notice that we could get the same 
amplitude if the model contains only the Higgs boson with
renormalized parameters:
\begin{equation}
L = \frac{1}{2}\left(\partial _{\mu }h \right)\left(\partial ^{\mu } h \right) - \frac{1}{2}m^2_r h^2 - \frac{\xi}{2}\frac{m^2_r}{\Lambda_r} h^3 + ...
\end{equation}
In other words, the radion contribution can mimic the deviation in
the triple Higgs coupling. This fact must be taken into account in
the investigation of $h^3$ coupling in the case of the radion
detection. However this is valid only for the processes in the
first order in the radion coupling constant, in more complicated
cases the difference between the radion and the Higgs boson can be
more significant and go beyond the simple replacement of the
Higgs potential parameters.

\section{Cancellations of additional to the Higgs-like contributions in associated Higgs boson-radion production}
\label{sec:Cancellation_associated}

In paper \cite{Boos:2014} it was shown  that all the additional
contributions as compared to the Higgs boson case are cancelled
out in the amplitudes of the single radion production processes. This
property follows from the structure of any massive fermion current
emitting the radion and an arbitrary number of any SM gauge
bosons. Now let us  show that the similar general property
takes place in the case of the associated Higgs boson-radion
production. Above  we have already demonstrated the explicit
cancellation by the example of the associated Higgs boson-radion
production in fermion-antifermion annihilation and in gluon
fusion. These were the processes with only two bosons ($rh$ or
$hh$) in the final state. For the general proof let us consider a
fermion current (or a fermion loop) with the emission of an arbitrary
number, say $N$, of SM bosons (vector gauge -- $V$ or Higgs --
$h$) with all possible permutations. $N_V$ stands for the number of gauge
bosons and $N_h$ -- for the number of Higgs bosons, $N_V+N_h=N$.
Now add  another Higgs boson (for $f\bar{f} \rightarrow
h_1,..., h_{N_h+1}, V_1,..., V_{N_V}$) or another radion (for
$f\bar{f} \rightarrow r, h_1,..., h_{N_h}, V_1,..., V_{N_V}$)  to
this current in all possible ways.

\begin{figure*}[!h!]
\begin{center}
\includegraphics[height=8cm]{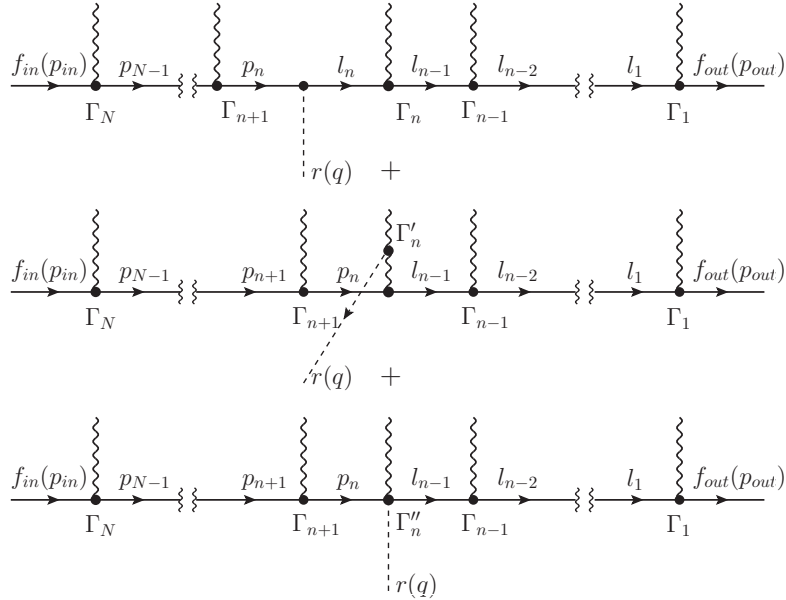}
\end{center}
\caption[]{Fermion current radiating the radion and $N$ SM Higgs or vector gauge bosons, $l_j \equiv p_j-q$.} \label{fig:diag_r} \vspace*{-2mm}
\end{figure*}

There are two possibilities of adding a Higgs boson : a) the
one emitted from the fermion line, and b) the one emitted from the
boson ($V$ or $h$) line. For the radion there are the same
options plus another one: c) the radion is emitted directly from
the four-point vertex with $V$ or $h$ boson
(Fig.\ref{fig:diag_r}).

Let us first set the notations for all the vertices. For
each vertex here all the lines are considered to be
incoming. We will use Latin indices ($a, b, c$) for a
parallel consideration of cases with gauge and Higgs bosons. In
the case of gauge bosons the Latin indices take values of vector
indices ($\alpha, \beta, \gamma, \mu$) and in  the case of
Higgs bosons they just reduce to the  sign $(h)$ for Higgs boson
vertices.

So, $\Gamma_a$ stands for the Lorentz part of the fermion-fermion-gauge boson vertex and for the fermion-fermion-Higgs boson vertex respectively
\begin{equation}
\Gamma_a =
  \begin{cases}
   \Gamma_\mu \\
   \Gamma_{(h)}
  \end{cases}
  =i
  \begin{cases}
   \gamma_\mu \left(a_f+b_f\gamma_5\right)\\
   -\frac{m_f}{v}
  \end{cases}
\end{equation}
In the same manner we denote the vertex with two gauge bosons and the Higgs boson and the triple Higgs vertex
\begin{equation}
\left[ \Gamma^\prime_{(h)} \right]^a_b =
  \begin{cases}
   \left[ \Gamma^\prime_{(h)} \right]^\alpha_\beta \\
   \Gamma^\prime_{(h)}
  \end{cases}
  =i
  \begin{cases}
   \frac{2m^2_V}{v}\delta^\alpha_\beta \\
   -\frac{3m^2_h}{v}
  \end{cases}
\end{equation}

Now for the radion, the fermion-fermion-radion vertex
$\Gamma_{(r)}$ is a function of the momenta of the incoming
fermions, say $p$ and $l$. It can be rewritten in terms of
the inverse propagators $S^{-1}$ \cite{Boos:2014} and takes the
following form
\begin{equation}
\begin{split}
\Gamma_{(r)} (p, l)= -\frac{i}{\Lambda_r} \left\{ \frac{3}{2} \left[ (\cancel{p}+m_f)-(\cancel{l}-m_f) \right]+m_f \right\}\\
= \frac{i}{\Lambda_r} \left\{ \frac{3}{2} \left[ S^{-1}(l) + S^{-1}(-p)\right] - m_f \right\}\\
=\frac{v}{\Lambda_r}\Gamma_{(h)} + \frac{i3}{2\Lambda_r}\left[ S^{-1}(l) + S^{-1}(-p) \right]
\end{split}
\label{Gamma_r}
\end{equation}
The notation $\left[ \Gamma^\prime_{(r)} \right]^a_b$
unifies the vertices for two gauge bosons and the radion
interaction (excluding the anomalies) and for two Higgs bosons and
the radion interaction respectively
\begin{equation}
\left[ \Gamma^\prime_{(r)} \right]^a_b=
\begin{cases}
   \left[ \Gamma^\prime_{(r)} \right]^\alpha_\beta \\
   \Gamma^\prime_{(r)}
  \end{cases}
=i
  \begin{cases}
   \frac{2m^2_V}{\Lambda_r} \delta^\alpha_\beta \\
   -\frac{2}{\Lambda_r} \left\{ (p \cdot k) +2m^2_h \right\}
  \end{cases}
\end{equation}
The term $\Gamma^{\prime\prime}_{(r),a}$ stands for gauge
boson-fermion-fermion-radion and Higgs
boson-fermion-fermion-radion four-point vertices
\begin{equation}
\Gamma^{\prime\prime}_{(r),a}=
\begin{cases}
   \left[ \Gamma^{\prime\prime}_{(r)} \right]_\mu \\
   \Gamma^{\prime\prime}_{(r)}
  \end{cases}
  =\frac{i}{\Lambda_r}
  \begin{cases}
   3\gamma_\mu \left(a_f+b_f\gamma_5\right) \\
   -\frac{4m_f}{v}
  \end{cases}
=\frac{1}{\Lambda_r}
  \begin{cases}
   3\Gamma_\mu \\
   4\Gamma_{(h)}
  \end{cases}
  =\frac{3}{\Lambda_r}\Gamma_a+\frac{1}{\Lambda_r}
  \begin{cases}
   0 \\
   \Gamma_{(h)}
  \end{cases}
\end{equation}

Now one can write down the contributions to the amplitudes in the following form
\begin{equation}
M[\varphi]_n=i^N \bar{f}(p_{out}) \left[ \prod_{j=1}^n \epsilon(k_j)^{a_j} \Gamma_{a_j} S(p_j-q) \right] \varphi(q) \Gamma_{(\varphi)} \left[ \prod_{j=n+1}^N S(p_{j-1}) \epsilon(k_j)^{a_j} \Gamma_{a_j} \right] f(p_{in})
\label{M_n}
\end{equation}
\begin{equation}
\begin{split}
M^\prime[\varphi]_n=i^N \bar{f}(p_{out}) \left[ \prod_{j=1}^{n-1} \epsilon(k_j)^{a_j} \Gamma_{a_j} S(p_j-q) \right] \left(\epsilon(k_n)^{a_n} \, \varphi(q) \, \left[ \Gamma^\prime_{(\varphi)}\right]^c_{a_n} G(k_n+q)^b_c \, \Gamma_b \right) \times \\
\left[ \prod_{j=n+1}^N S(p_{j-1}) \epsilon(k_j)^{a_j} \Gamma_{a_j} \right] f(p_{in})
\end{split}
\label{M_n_prime}
\end{equation}
\begin{equation}
\begin{split}
M^{\prime\prime}[\varphi]_n=i^{N-1} \bar{f}(p_{out}) \left[ \prod_{j=1}^{n-1} \epsilon(k_j)^{a_j} \Gamma_{a_j} S(p_j-q) \right] \left(\epsilon(k_n)^{a_n} \, r(q) \, \left[ \Gamma^{\prime\prime}_{(r)}\right]_{a_n}\right) \times \\
\left[ \prod_{j=n+1}^N S(p_{j-1}) \epsilon(k_j)^{a_j} \Gamma_{a_j} \right] f(p_{in})
\end{split}
\label{M_n_2prime}
\end{equation}
where $\varphi$ is either $h$ or $r$ in (\ref{M_n}), ({\ref{M_n_prime}) and only $r$ in ({\ref{M_n_2prime});
$\epsilon(k_j)^{a_j}$ is either $\epsilon(k_j)^{\mu_j}$ (for the gauge boson) or $h(k_j)$ (for the Higgs boson); $G(k_n+q)^b_c$ is either $G(k_n+q)^\beta_\gamma$ (the gauge boson propagator) or $G(k_n+q)$ (the Higgs boson propagator);
$$p_{n-1}=p_n-k_n, \quad p_{in}=p_N, \quad p_{out}=p_0-q.$$

The number $n$ runs from $1$ to $N$ ($n=1,2,...,N$). In the
case of real initial and final fermions one should take into
account another amplitude
\begin{equation}
M[\varphi]_0=i^N \left[ \prod_{j=1}^N \epsilon(k_j)^{a_j} \right] \varphi(q) \bar{f}(p_{out}) \Gamma_{(\varphi)} \left[ \prod_{j=1}^N S(p_{j-1}) \Gamma_{a_j} \right] f(p_{in})
\label{M_0}
\end{equation}
If the added particle is the Higgs boson, the amplitude takes the form
\begin{equation}
M[h]_0=i^N \bar{f}(p_{out}) h(q) \Gamma_{(h)} \left[ \prod_{j=1}^N S(p_{j-1}) \epsilon(k_j)^{a_j} \Gamma_{a_j} \right] f(p_{in})
\label{M_0_h}
\end{equation}
If the added particle is the radion, one can write the following
 amplitude
\begin{equation}
\begin{split}
M[r]_0=\frac{v}{\Lambda_r} i^N \left[ \prod_{j=1}^N \epsilon(k_j)^{a_j} \right] r(q) \bar{f}(p_{out}) \Gamma_{(h)} \left[ \prod_{j=1}^N S(p_{j-1}) \Gamma_{a_j} \right] f(p_{in})\\
+\frac{i3}{2\Lambda_r} i^N \left[ \prod_{j=1}^N \epsilon(k_j)^{a_j} \right] r(q) \bar{f}(p_{out}) S^{-1}(p_0)
\left[ \prod_{j=1}^N S(p_{j-1}) \Gamma_{a_j} \right] f(p_{in})\\
=\frac{v}{\Lambda_r} M[h]_0 +\frac{i3}{2\Lambda_r} i^N \left[ \prod_{j=1}^N \epsilon(k_j)^{a_j} \right] r(q) \bar{f}(p_{out}) \left[ \prod_{j=1}^N \Gamma_{a_j} S(p_{j}) \right] \Gamma_{a_N} f(p_{in})
\end{split}
\label{M_0_r}
\end{equation}
where we used (\ref{Gamma_r}) and the equation of motion
$\bar{f}(p_{out}) S^{-1}(p_{out})=0$ ($p_{out}$ is the
outgoing momentum).

Now let us take any number $n$, $n=1,...,N$. In the case of
 adding a Higgs boson  the sum of all amplitudes for the
chosen $n$ is
\begin{equation}
\begin{split}
M[h]_n+M^\prime[h]_n=i^N \left[ \prod_{j=1}^N \epsilon(k_j)^{a_j} \right] h(q) \bar{f}(p_{out}) \left[ \prod_{j=1}^{n-1} \Gamma_{a_j} S(p_j-q) \right]\\ \times
\left\{ \Gamma_{a_n} S(p_n-q) \Gamma_{(h)} + \left[\Gamma^\prime_{(h)} \right]^c_{a_n} G(k+q)^b_c \, \Gamma_b \right\} \left[ \prod_{j=n+1}^N S(p_{j-1}) \Gamma_{a_j} \right] f(p_{in})
\end{split}
\label{M_n_h_sum}
\end{equation}
For the case of adding a radion  the sum has the following
form
\begin{equation}
\begin{split}
M[r]_n+M^\prime[r]_n+M^{\prime\prime}[r]_n=i^N \left[ \prod_{j=1}^N \epsilon(k_j)^{a_j} \right] r(q) \bar{f}(p_{out}) \left[ \prod_{j=1}^{n-1} \Gamma_{a_j} S(p_j-q) \right]\\ \times
\left\{ \Gamma_{a_n} S(p_n-q) \Gamma_{(r)} + \left[\Gamma^\prime_{(r)} \right]^c_{a_n} G(k+q)^b_c \, \Gamma_b - i\left[\Gamma^{\prime\prime}_{(r)} \right]_{a_n} \right\} \left[ \prod_{j=n+1}^N S(p_{j-1}) \Gamma_{a_j} \right] f(p_{in})
\end{split}
\label{M_n_r_sum}
\end{equation}

One can calculate the part in curly brackets in (\ref{M_n_r_sum}) and compare it with that in (\ref{M_n_h_sum})
\begin{equation}
\begin{split}
\Gamma_{a_n} S(p_n-q) \Gamma_{(r)} + \left[\Gamma^\prime_{(r)} \right]^c_{a_n} G(k+q)^b_c \, \Gamma_b - i\left[\Gamma^{\prime\prime}_{(r)} \right]_{a_n}=\\
\frac{v}{\Lambda_r} \left( \Gamma_{a_n} S(p_n-q) \Gamma_{(h)} +
\begin{Bmatrix}
    1 \\
    1+\xi
  \end{Bmatrix}
\left[\Gamma^\prime_{(h)} \right]^c_{a_n} G(k+q)^b_c \, \Gamma_b \right)\\
+ \frac{i3}{2\Lambda_r} \Gamma_{a_n} (S(p_n-q) S^{-1}(p_n) -1)
\end{split}
\label{curly_r}
\end{equation}

Substituting the result (\ref{curly_r}) into expression
(\ref{M_n_r_sum}) and opening the brackets one gets
\begin{equation}
\begin{split}
M[r]_n+M^\prime[r]_n+M^{\prime\prime}[r]_n=i^N \left[ \prod_{j=1}^N \epsilon(k_j)^{a_j} \right] r(q) \bar{f}(p_{out}) \left[ \prod_{j=1}^{n-1} \Gamma_{a_j} S(p_j-q) \right]\\ \times
\frac{v}{\Lambda_r} \left( \Gamma_{a_n} S(p_n-q) \Gamma_{(h)} +
\begin{Bmatrix}
    1 \\
    1+\xi
  \end{Bmatrix}
\left[\Gamma^\prime_{(h)} \right]^c_{a_n} G(k+q)^b_c \, \Gamma_b \right)
\left[ \prod_{j=n+1}^N S(p_{j-1}) \Gamma_{a_j} \right] f(p_{in})\\
+ i^N \left[ \prod_{j=1}^N \epsilon(k_j)^{a_j} \right] r(q) \bar{f}(p_{out}) \left[ \prod_{j=1}^{n-1} \Gamma_{a_j} S(p_j-q) \right]\\ \times
\frac{i3}{2\Lambda_r} \Gamma_{a_n} (S(p_n-q) S^{-1}(p_n) -1)
\left[ \prod_{j=n+1}^N S(p_{j-1}) \Gamma_{a_j} \right] f(p_{in})
\end{split}
\label{M_n_r_sum1}
\end{equation}
Here the first term  has almost the same form as in the
case of the Higgs boson (the absolute similarity would take place
if $\xi=0$ and $v=\Lambda_r$). In the second term one can open the
brackets and get
\begin{equation}
\begin{split}
i^N \left[ \prod_{j=1}^N \epsilon(k_j)^{a_j} \right] r(q) \bar{f}(p_{out}) \left[ \prod_{j=1}^{n-1} \Gamma_{a_j} S(p_j-q) \right]
\frac{i3}{2\Lambda_r} \Gamma_{a_n} (S(p_n-q) S^{-1}(p_n) -1)\\
\times \left[ \prod_{j=n+1}^N S(p_{j-1}) \Gamma_{a_j} \right] f(p_{in})\\
=\frac{i3}{2\Lambda_r} i^N \left[ \prod_{j=1}^N \epsilon(k_j)^{a_j} \right] r(q) \bar{f}(p_{out}) \left[ \prod_{j=1}^{n} \Gamma_{a_j} S(p_j-q) \right] \Gamma_{a_{n+1}} \left[ \prod_{j=n+2}^N S(p_{j-1}) \Gamma_{a_j} \right] f(p_{in})\\
-\frac{i3}{2\Lambda_r} i^N \left[ \prod_{j=1}^N \epsilon(k_j)^{a_j} \right] r(q) \bar{f}(p_{out}) \left[ \prod_{j=1}^{n-1} \Gamma_{a_j} S(p_j-q) \right] \Gamma_{a_n} \left[ \prod_{j=n+1}^N S(p_{j-1}) \Gamma_{a_j} \right] f(p_{in})\\
=U_{n+1}-U_n
\end{split}
\label{M_n_r_sum2}
\end{equation}
where
\begin{equation}
U_n=\frac{i3}{2\Lambda_r} i^N \left[ \prod_{j=1}^N \epsilon(k_j)^{a_j} \right] r(q) \bar{f}(p_{out}) \left[ \prod_{j=1}^{n-1} \Gamma_{a_j} S(p_j-q) \right] \Gamma_{a_n} \left[ \prod_{j=n+1}^N S(p_{j-1}) \Gamma_{a_j} \right] f(p_{in})
\label{U_n}
\end{equation}
for $n<N$.

It is easy to write $U_1$ in the following form
\begin{equation}
U_1=\frac{i3}{2\Lambda_r} i^N \left[ \prod_{j=1}^N \epsilon(k_j)^{a_j} \right] r(q) \bar{f}(p_{out}) \left[ \prod_{j=1}^{N-1} \Gamma_{a_j} S(p_j) \right] \Gamma_{a_N} f(p_{in})
\label{U_n1}
\end{equation}

Getting back to the expression for $M[r]_0$ (\ref{M_0_r}) one finds
\begin{equation}
M[r]_0=\frac{v}{\Lambda_r} M[h]_0 + U_1
\end{equation}

For $n=N$ one must separately consider the case of real initial
and final fermions and the case of a fermion loop. For the
first case the equation of motion $S^{-1}(p_N)f(p_{in})=0$ is valid,
thus
\begin{equation}
\begin{split}
i^N \left[ \prod_{j=1}^N \epsilon(k_j)^{a_j} \right] r(q) \bar{f}(p_{out}) \left[ \prod_{j=1}^{N-1} \Gamma_{a_j} S(p_j-q) \right]
\frac{i3}{2\Lambda_r} \Gamma_{a_N} (S(p_N-q) S^{-1}(p_N) -1)\\
=-\frac{i3}{2\Lambda_r} i^N \left[ \prod_{j=1}^N \epsilon(k_j)^{a_j} \right] r(q) \bar{f}(p_{out}) \left[ \prod_{j=1}^{N-1} \Gamma_{a_j} S(p_j-q) \right] \Gamma_{a_N} f(p_{in})\\
=-U_N
\end{split}
\end{equation}

Finally for the case of real initial and final fermions we have
\begin{equation}
\begin{split}
M[r]_0+ \sum_{n=1}^N (M[r]_n + M^\prime[r]_n + M^{\prime\prime}[r]_n)=
\frac{v}{\Lambda_r} M[h]_0
+\frac{v}{\Lambda_r} \sum_{n=1}^N \left( M[h]_n +
\begin{Bmatrix}
    1 \\
    1+\xi
  \end{Bmatrix}
M^\prime[h]_n
\right)\\
+U_1 +(U_2-U_1) +(U_3-U_2) + \ldots +(U_N-U_{N-1}) -U_N=\\
\frac{v}{\Lambda_r} M[h]_0
+\frac{v}{\Lambda_r} \sum_{n=1}^N \left( M[h]_n +
\begin{Bmatrix}
    1 \\
    1+\xi
  \end{Bmatrix}
M^\prime[h]_n
\right)
\end{split}
\label{M_n_r_sum3}
\end{equation}

In the case of the fermion loop ($p_{in}=p_{out}$) one can move $\bar{f}(p_{out})$ by
cyclic permutations to the end of the
matrix product, which leaves the trace invariant, so $f(p_{in})\bar{f}(p_{out})=S(p_{in})=S(p_{out})$ and therefore $S^{-1}(p_{N})f(p_{in})\bar{f}(p_{out})=1$.
\begin{equation}
\begin{split}
i^N \left[ \prod_{j=1}^N \epsilon(k_j)^{a_j} \right] r(q) \bar{f}(p_{out}) \left[ \prod_{j=1}^{N-1} \Gamma_{a_j} S(p_j-q) \right]
\frac{i3}{2\Lambda_r} \Gamma_{a_N} (S(p_N-q) S^{-1}(p_N) -1)f(p_{in})\\
=\frac{i3}{2\Lambda_r} i^N \left[ \prod_{j=1}^N \epsilon(k_j)^{a_j} \right] r(q) \left[ \prod_{j=1}^N \Gamma_{a_j} S(p_j-q) \right]-U_N
\end{split}
\label{terms4loop}
\end{equation}

It is easy to check that the last but one term in
(\ref{terms4loop}) is equal to $U_1$. Indeed, it can be shown by
means of the same trick: moving $f(p_{in})$ to the beginning and
shifting the loop momentum by the value $q$. Thus, just as in the
case of real fermions we get (\ref{M_n_r_sum3}).


\section{Conclusions}
\label{sec:Conclusions} In the current work we have discussed the
 Higgs boson-radion similarity in their associated production
processes. First, the associated Higgs boson-radion production was
considered in two examples -- in fermion-antifermion
annihilation and gluon fusion. In both cases we have shown
explicitly the Higgs boson-radion similarity up to the
replacement of the masses and the denominators of the coupling
constants and a rescaling of the triple Higgs coupling.
Next, the general proof of this property was provided for the case
of the radion production in association with an arbitrary number
of the SM gauge or Higgs bosons. It was found that the Higgs
boson-radion similarity in the considered types of processes does
not allow us to distinguish a model with the radion
and the Higgs boson from a  model without the radion but
with the Higgs boson with modified parameters. In particular, the
radion contribution can mimic the deviation in the triple Higgs
coupling. This fact must be taken into account in the
investigation of $h^3$ coupling in the case of the radion
detection.

Of course there exists the well-known difference between the radion
and the Higgs boson because of the presence of the radion
anomalous interaction. In addition to the enhancement of the
radion decay modes to two gluons and to two photons the anomalous
radion-gluon-gluon interaction contributes differently to the
associated Higgs boson-radion and to the Higgs pair production.
In the latter case the Higgs boson pair production may occur via
the radion decay. The corresponding diagram does not
participate in the  cancellations and turns to a diagram of the
same $\Lambda_r^{-1}$ order in the case of the resonant Higgs boson
production ($m_r>2m_h$) while in the case of the non-resonant Higgs boson
production ($m_r<2m_h$) this diagram is of the next order
($\Lambda_r^{-2}$). In fact, the anomalous
radion-gluon-gluon interaction  gives the leading
contribution to the Higgs pair production via the radion decay.

The radion pair production is not considered in the current work being a model dependent
and complicated study where the next orders ought to be taken into account.

It is important to mention that the considered property is valid not only for the radion in the brane-world models with two branes but it can also take place in scalar-tensor gravity theories (for example, the Brans-Dicke theory) or theories involving dilaton where the scalar field interacts with the trace of the energy-momentum tensor of matter.

\section{Acknowledgments}
The work was supported by   grant 14-12-00363 of Russian Science
Foundation. The authors are grateful to V.~Bunichev, M.~Smolyakov,
and I.~Volobuev for useful discussions and critical remarks.


\bibliographystyle{model1-num-names}



\end{document}